\documentclass[11pt,twoside]{article}
\usepackage{asp2006}
\usepackage{graphics}
\usepackage{hyperref}
\usepackage{graphicx}
\usepackage{lscape}
\newcommand{\XMM}{{\em XMM-Newton }}
\newcommand{\Ch}{{\em Chandra }}
\newcommand{\ot}{[O~{\sc iii}] }

\def\edcomment#1{\iffalse\marginpar{\raggedright\sl#1\/}\else\relax\fi}
\reversemarginpar

\begin{document}
\title{Searching for AGN Outflows: Spatially Resolved Chandra HETG Spectroscopy of the NLR Ionization Cone in NGC 1068}

\author{Daniel A. Evans\altaffilmark{1}, Patrick M. Ogle\altaffilmark{2}, Herman L. Marshall\altaffilmark{1}, Mike A. Nowak\altaffilmark{1}, Stefano Bianchi\altaffilmark{3}, Matteo Guainazzi\altaffilmark{4}, Anna Lia Longinotti\altaffilmark{1}, Dan Dewey\altaffilmark{1}, Norbert S. Schulz\altaffilmark{1}, Mike S. Noble\altaffilmark{1}, John Houck\altaffilmark{1}, and Claude R. Canizares\altaffilmark{1}}

\altaffiltext{1}{MIT Kavli Institute for Astrophysics and Space Research, 77 Massachusetts Avenue, Cambridge, MA 02139, USA}
\altaffiltext{2}{Spitzer Science Center, California Institute of Technology, Mail Code 220-6, Pasadena, CA 91125, USA}
\altaffiltext{3}{Dipartimento di Fisica, Universita degli Studi Roma Tre, via della Vasca Navale 84, I-00146 Roma, Italy}
\altaffiltext{4}{European Space Astronomy Centre of ESA, P.O.Box 78, Villanueva de la Canada, E-28691 Madrid, Spain}

\begin{abstract}
We present initial results from a new 440-ks Chandra HETG GTO observation of the canonical Seyfert 2 galaxy NGC 1068. The proximity of NGC 1068, together with Chandra's superb spatial and spectral resolution, allow an unprecedented view of its nucleus and circumnuclear NLR. We perform the first spatially resolved high-resolution X-ray spectroscopy of the `ionization cone' in any AGN, and use the sensitive line diagnostics offered by the HETG to measure the ionization state, density, and temperature at discrete points along the ionized NLR. We argue that the NLR takes the form of outflowing photoionized gas, rather than gas that has been collisionally ionized by the small-scale radio jet in NGC 1068. We investigate evidence for any velocity gradients in the outflow, and describe our next steps in modeling the spatially resolved spectra as a function of distance from the nucleus. \end{abstract}

\vspace{-0.5cm}
\section{Overview: The Role of AGN Outflows in Galaxy Evolution}

Outflows and feedback from AGN are widely invoked as the key mediator between the co-evolution of black holes and their host galaxies over cosmic time. It is now well understood from large optical surveys that galaxies evolve through mergers from blue, star-forming spirals (the `blue cloud') whose black holes accrete at or near their Eddington limits, through a transition region (the `green valley'), to so-called `red and dead' ellipticals (the `red sequence'), which instead are described by markedly less black-hole growth. The importance of outflows in this evolution has now taken center stage; yet before we can successfully incorporate AGN feedback into numerical simulations of galaxy growth, a number of key questions need to be answered from observations:

\begin{itemize}
\item Can AGN actually deliver enough power to their environments to alter the evolution of their host galaxies in a meaningful way?
\item On what spatial scales does this occur?
\item Where are the outflows in the first place?
\end{itemize}

Several steps have recently been taken to address these issues. It is now apparent from multiwavelength surveys that AGN selected in different wavelengths tend to populate different regions of the galaxy color-magnitude diagram. For example, \cite{hic09} showed that IR-selected AGN are characterized by high Eddington ratios and tend to inhabit the `blue cloud'; X-ray AGN have intermediate-to-high ratios and lie in the `green valley'; and radio AGN have low Eddington fractions and populate the `red sequence'. The relative paucity of galaxies in the `green valley', together with the prevalence of X-ray AGN in this region suggest that the kpc-scale, ionized Narrow-Line Regions (NLR) in such AGN are ideal places to search for outflows, together with their kinematic and ionizing effect on their galaxy-scale environments. 

\section{X-Ray Observations of Outflows in AGN: Introducing NGC 1068}

Observations of Type 1 Seyfert galaxies with the high-resolution X-ray gratings instruments on board \Ch and \XMM have yielded strong evidence for outflows seen in absorption against the nuclear continuum, with velocities of hundreds to thousands of km~s$^{-1}$ (e.g., \citealt{ree09,ste09}). However, owing to the face-on orientation of these systems with respect to the observer, there is little information on the {\it spatial extent} of these outflows. Type 2 Seyfert galaxies, on the other hand, which have edge-on inclinations, do not suffer from this orientation issue; yet since only a small percentage of the nuclear flux is scattered into the line of sight by a medium external to the AGN, the signal-to-noise of any outflow seen this time in {\it emission} tends to be poor and is dominated by the nucleus, rather than the off-nuclear gas.

There is one AGN, the prototypical Seyfert 2 galaxy NGC~1068, that is sufficiently near ($z$=0.003793; 1$''$=80~pc) and bright that we can, in principle, perform {\it spatially resolved, high-resolution gratings spectroscopy} of both the nucleus {\it and} any off-nuclear gas. NGC 1068 has a $10^{7}$ $M_\odot$ black hole that is accreting at or close to its Eddington limit (e.g., \citealt{kis02}), making it an excellent source to examine the role of AGN outflows on galaxy-scale gas on a galaxy in the `green valley'. NGC~1068 also shows evidence of a kpc-scale radio jet (\citealt{vdh82}), meaning that we can investigate the competing roles of collisional ionization from the radio ejecta and photoionization from the AGN radiation field. We can also compare these effects between Seyfert galaxies and radio-loud AGN, such as the well-studied radio galaxy 3C\,33 (\citealt{kra07,tor09}).

We have recently obtained a 440-ks \Ch HETG GTO observation of NGC~1068 (PI Canizares), and present our initial results here. In particular, we use:

\begin{enumerate}
\item Multiwavelength imaging of kpc-scale circumnuclear gas; and
\item High-resolution \Ch HETG spectroscopy
\end{enumerate}

\noindent to study the NLR in NGC 1068, and in doing so determine the physical conditions and kinematics of any AGN-induced outflow.

\section{Multiwavelength Imaging of NGC 1068}

In Figure 1, we show a tri-color X-ray image of NGC 1068 (left), together with a multiwavelength (X-ray, \ot and radio) overlay (right). It is clear from the left panel that the hard X-ray emission tends to principally be concentrated in the nucleus itself, though there is evidence that it extends $\sim$1~kpc from the AGN in a biconical distribution. The soft X-rays are much more spatially elongated, and extend several kpc from the nucleus in either direction. The multiwavelength image (right panel) demonstrates the complex morphology between the X-ray and \ot NLR, and the kpc-scale radio jet. Although there is evidence for jet-gas interactions, it is evident that the soft X-ray emission and \ot lie several kpc {\it beyond} the radio ejecta, which suggests that shock-heating of gas by the jet does not dominate the energetics of the source. However, it is only by performing high-resolution spectroscopy that we can distinguish between collisional ionization from the jet and photoionization from the AGN radiation field, and test if the extended X-ray gas is ambient, or instead takes the form of an outflow.

\begin{figure}
\centering
\includegraphics[scale=0.35]{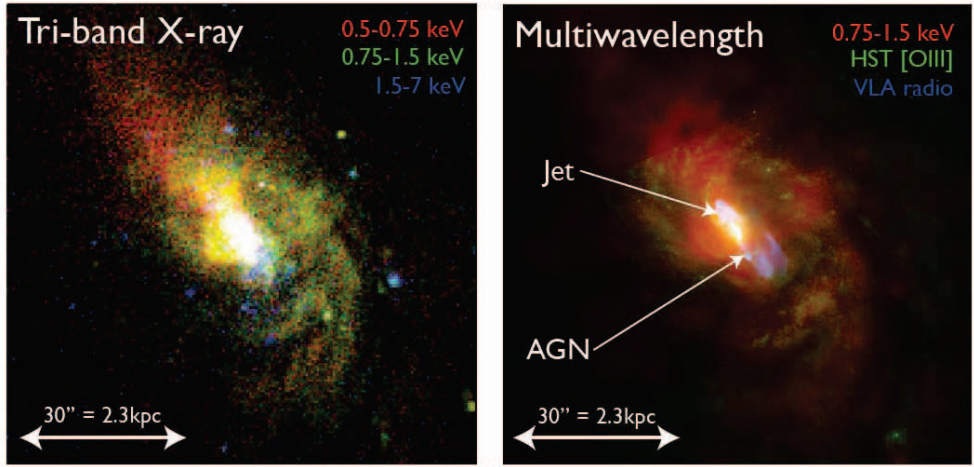}
\label{multiwavelength}
\caption{({\it left}): HETG zero-order image of NGC~1068. Shown are events in the 0.5--0.75 keV (red), 0.75--1.5 keV (green), and 1.5--7 keV (blue) energy bands. ({\it right}): Multiwavelength composite of NGC 1068, in 0.75--1.5 keV X-rays (red), HST \ot (green), and VLA radio (blue) bands.}
\end{figure}

\section{HETG Spectroscopy of NGC 1068}

\begin{figure}
\centering
\includegraphics[angle=270, scale=0.7]{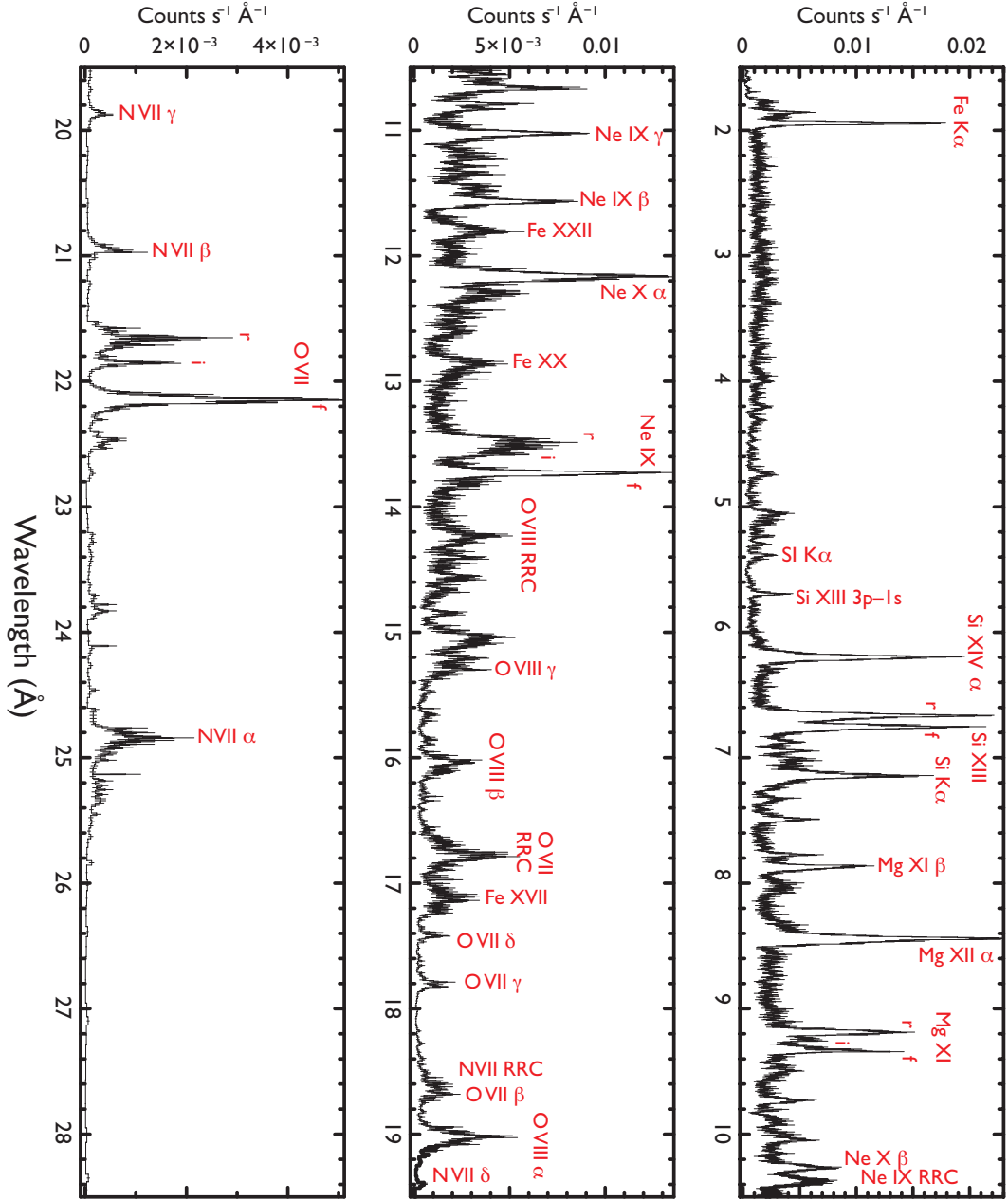}
\caption{Co-added \Ch HETG MEG and HEG gratings spectrum of NGC 1068 from the entire 440-ks data set. Shown are the principal transitions from H- and He-like species, together with RRCs.}
\label{3panel}
\end{figure}

\begin{figure}
\includegraphics[width=14cm]{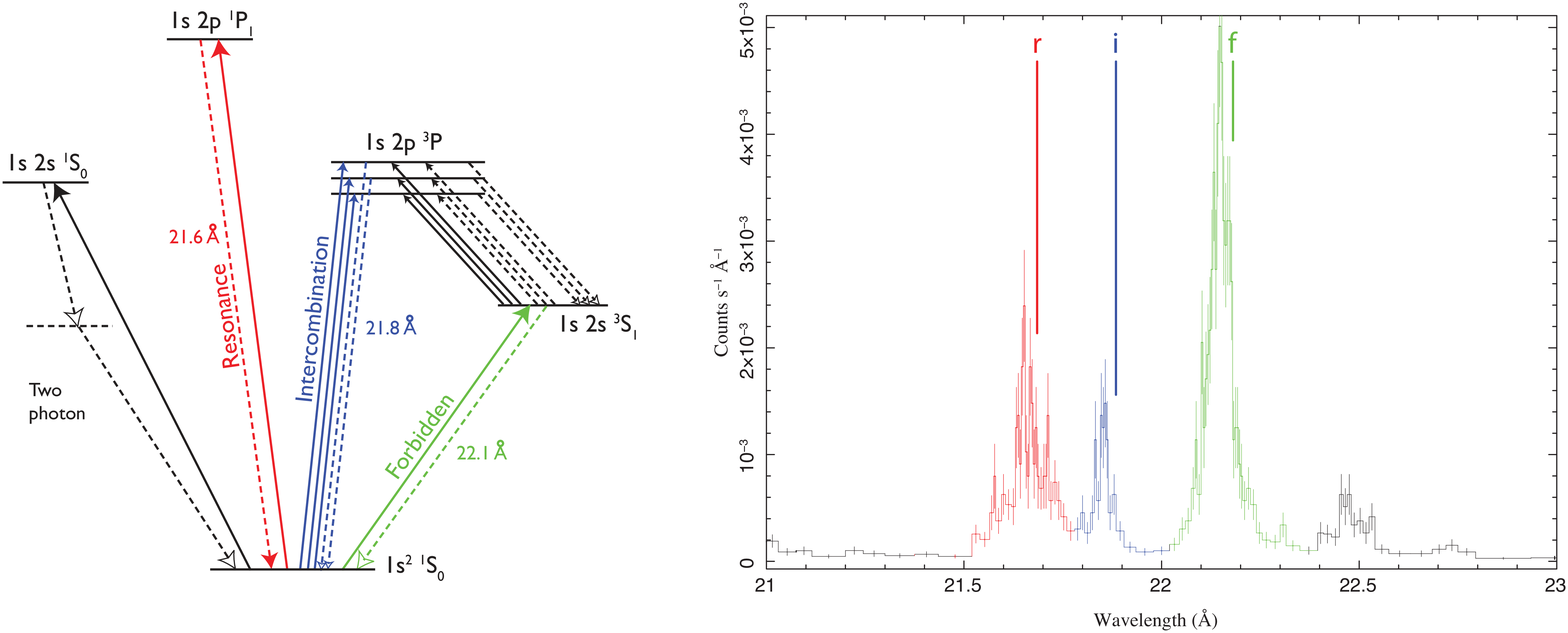}
\label{rif}
\caption{({\it left}): Energy level diagram for a two-electron (He-like ion), showing the radiative decays from the $n=2$ to $n=1$ level that form the resonance (red), intercombination (blue), and forbidden (green) transitions. The wavelengths apply to the case of O~{\sc vii}. ({\it right}): The O~{\sc vii} triplet observed in NGC 1068. Also plotted are the observer's frame wavelengths of the resonance, intercombination, and forbidden lines. The observed blueshifts from these wavelengths indicate that the gas is outflowing.}
\end{figure}

In Figure 2 we show the co-added \Ch MEG and HEG spectrum of the nucleus of NGC 1068 from the entire 440-ks observation. We plot the principal transitions of H- and He-like species, together with radiative recombination continuua (RRCs). Many of these transitions have already been identified by \cite{kin02} and \cite{ogle03}. In Figure 3, we show an energy level diagram for a two-electron (He-like ion), showing the radiative decays from the $n=2$ to $n=1$ level that form the resonance (red), intercombination (blue), and forbidden (green) transitions (left panel). The wavelengths apply to the case of O~{\sc vii}. In the right panel, we show the O~{\sc vii} triplet observed in NGC 1068. Also plotted are the observer's frame wavelengths of the resonance, intercombination, and forbidden lines. The observed blueshifts from these wavelengths indicate that the gas is outflowing, with velocities of several hundred km~s$^{-1}$. The detection of narrow RRCs and the ratios of the resonance, intercombination, and forbidden lines strongly suggest that the gas is {\it photoionized} by the AGN radiation field and takes the form of an {\it outflow}.

\section{Next Steps: Spatially Resolved HETG Spectroscopy}

In Figure 4, we show the an video of the \Ch HETG spectrum of NGC~1068 between -1 and +1 kpc, in 1$''$ (80~pc) steps. There is sufficient signal-to-noise to enable us to perform detailed photoionization modeling of the off-nuclear gas along this ionization cone. Our preliminary work indicates that {\it outflows of several hundred km~s$^{-1}$ are detected out to 1 kpc from the AGN}. From our ongoing photoionization modeling, we will be able to calculate the mass outflow rate and power deposited by the photoionized outflow into the galaxy-scale gas, providing key constraints need to answer the long standing question about the role of AGN outflows on galaxy evolution.

\begin{figure}[th]
\centering
\href{http://www.vimeo.com/7084990}{\includegraphics[width=14cm]{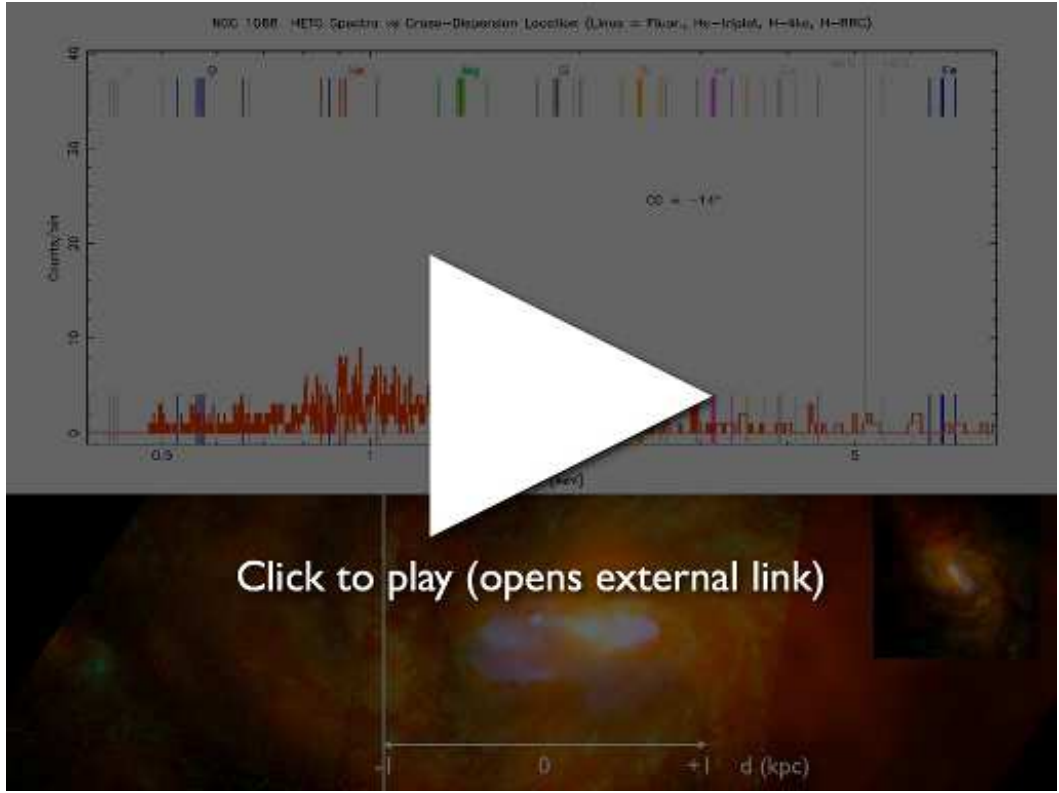}}
\caption{Video of the \Ch HETG spectrum of NGC 1068, ranging between -1 and +1 kpc in steps of 1$''$ (80~pc). \href{http://www.vimeo.com/7084990}{Click here to launch}.}
\label{offaxisspectrum}
\end{figure}


\end{document}